Title: Cluster analysis of stocks using price movements of high frequency data from National Stock Exchange.

Abstract:

This paper aims to develop new techniques to describe joint behavior of stocks, beyond regression and correlation. For example, we want to identify the clusters of the stocks that move together. Our work is based on applying Kernel Principal Component Analysis(KPCA) and Functional Principal Component Analysis(FPCA) to high frequency data from NSE. Since we dealt with high frequency data with a tick size of 30 seconds, FPCA seems to be an ideal choice. FPCA is a functional variant of PCA where each sample point is considered to be a function in Hilbert space $L^2$. On the other hand, KPCA is an extension of PCA using kernel methods. Results obtained from FPCA and Gaussian Kernel PCA seems to be in synergy but with a lag. There were two prominent clusters that showed up in our analysis, one corresponding to the banking sector and another corresponding to the IT sector. The other smaller clusters were seen from the automobile industry and the energy sector. IT sector was seen interacting with these small clusters. The learning gained from these interactions is substantial as one can use it significantly to develop trading strategies for intraday traders.

Authors: Charu Sharma, Ph.D. scholar, Assistant Professor, School of Natural Sciences, Shiv Nadar University, UP; Prof. Amber Habib, Professor, School of Natural Sciences, Shiv Nadar University, UP; Prof. Sunil Bowry, Professor, School of Management and Entrepreneurship, Shiv Nadar University, UP.

Address: Charu Sharma, A111D, Shiv Nadar University, NH91, Tehsil Dadri, Gautam Buddha Nagar, Uttar Pradesh – 201314. Phone No: +91-9911750311, Email: charu.sharma@snu.edu.in

I. Introduction

According to a report by a *Brookernotes*, a UK based company, out of all the people who trade stocks online, about one-third of them are from Asia. In fact, out of the 3.2 million traders in Asia, 570,000 of them are based in India. In view of this, understanding the interactions amongst the stocks at a tick by tick level is a need of the hour. Among various factors, which influence the change in the stock prices, change in the prices of other stocks is one of the major influential factor. Over the years. researchers have used techniques like regression analysis and correlation to understand the co-movements of the stocks but that too on daily rate of return. In this paper we have tried to exploit the interactions amongst the stocks on tick by tick level where each tick is a 30 sec mark. The techniques used by us are extensions of a well-known classification technique called as Principal Component Analysis. Also when the sample points in the working dataset can be regarded as functions, then instead of using usual PCA for classification, one can think of using functional analogue of PCA called as functional PCA. Last two decades has seen tremendous development in the field of functional data analysis, a branch of statistics that deals with the data that consider each sample point as functions. Over the past two decades, Ramsay and Silverman[2-7] has shown many real world applications in the field of FDA. Since we worked with high frequency data, we can treat sample points as functions instead of a discrete set of values and thus FPCA looked a good choice for classifying the stocks. The second technique that has been used in this paper, is Kernel based principal component analysis (KPCA). This method is used to exploit the nonlinearity of the data if any. The data set is moved to a higher dimensional space where the new sets of points obey the linearity and thus PCA can be performed on this new set.

II. PCA, FPCA, KPCA

Principal Component analysis was introduced in early 20th century by Karl Pearson. PCA is a data reduction and classification technique. Under this technique, if we have n sample points with k features (usually k>n) then, we aim to find the basis for the subspace corresponding to the linear span of these n sample points in feature space $\mathbb{R}^k$. Clearly the dimension of this subspace will be less than or equal to n. Also we wish to place the basis elements in an order such that, the first basis element is the major factor that brings out differences amongst the sample points, the second basis element is the next major factor and so on. While aiming to find such a basis, it turns out that the basis elements are nothing but the eigenvectors of the covariance matrix in the feature space. Thus one carry out SVD of the covariance matrix to find its eigenvectors and corresponding eigenvalues. These basis elements are known as principal components.

Now in case when the features can be treated as a continuum, for example a quantity been measured at a regular interval of time be it every second or every minute, then the data can be regarded as a smooth function instead of discrete. In this case, our n sample points can be treated as n functions and if these functions are assumed to be in $\mathcal{L}^2$, then one again finds the basis of the linear span of these n sample points in $\mathcal{L}^2$. The basis elements of course must satisfy the order in the same sense as PCA. This variant of PCA is called FPCA. In our case since we are dealing with high frequency data  data picked at every 30 seconds; thus FPCA seems reasonable to use.

Once we get the basis, we express our data as linear combination of the principal components. We then use clustering algorithms like k-means clustering to cluster the data into different groups. K-means clustering however at times fails to consider nonlinearity of the data, if present. Thus the next method we tried is the nonlinear extension of PCA called as Kernel principal component analysis or KPCA. Here the idea is if the data is not linearly separable in $d < n$ dimensions, it however can almost always be linearly separable in higher dimensions. One defines a map $\phi: R^d \to R^N, N > d$ such that our data under this map is linearly separable in $R^N$. We have carried out analysis of the data by applying both FPCA and KPCA with Gaussian kernels and summarized the results obtained.

III. Data Description

We picked tick by tick data for the year 2014, from National Stock Exchange. We started with the stocks listed in CNX100 index, for that year. Initially all the 100 stocks listed in CNX100 index were picked but during the course of analysis 11 stocks were dropped due to insufficient data values or missing data. CNX100 index, consist of the Nifty50 and the CNX Nifty Junior stocks. Table 1 gives the detail of the composition. Also market opens at 9 o'clock in the morning and is functional till 4 PM, but the active trading sessions occurs between 9:30 till 3:30. Considering this, we have taken data between 9:30AM till 3:30PM, 6 hours a day, in our analysis. Every 30second is considered a tick, and thus in each day we have 720 tick points for each stock. For each stock, volume weighted average price (VWAP) per 30 seconds is calculated and used for further analysis.

| Industry Type | Number of stocks in CNX100 |
|---|---|
| INDUSTRIAL MANUFACTURING | 6 |
| CEMENT & CEMENT PRODUCTS | 4 |
| SERVICES | 2 |
| AUTOMOBILE | 10 |
| CONSUMER GOODS | 14 |
| PHARMA | 10 |
| FINANCIAL SERVICES | 14 |
| ENERGY | 10 |
| METALS | 6 |
| TELECOM | 3 |
| CONSTRUCTION | 2 |
| CHEMICALS | 1 |
| IT | 6 |

Table 1: Composition of index CNX100, 2014

| | |
|---|---|
| Number of stocks considered | 89 |
| Number of active trading days in 2014 | 229 |
| Number of daily ticks for each stock | 720 |
| Magnitude of data | 1,46,74,320 |

Table 2: Summary of the data

IV. Methodology and Analysis

In the past, many researchers have used correlation coefficient to understand the network amongst the stocks. We started our analysis with the same. Two stocks are taken at a time and Spearman rank correlation coefficients for each 3916 such pairs are calculated for every working day in 2014. Each day consists of 720 ticks. A sample of these correlation coefficients is given in figure 1. Figure 1 gives the correlation coefficient between each pair of stocks for the first trading day of each month. Also table 3 summarizes Figure 1. For most of these pairs, the correlation coefficient is observed to be less than 0.5, in fact they are as low as 0.2. One key observation of this analysis is that 8 out of 12 times, the maximum value of correlation coefficient, though very small, is seen to occur between PNB and Bank of Baroda. Two state owned multinational banks are seen moving hand in hand even at a 30 sec tick scale. We further investigate this by running k-means algorithm every day for 229 days to form clusters of these 89 stocks by using distance metric as $d(S_1, S_2) = 1 - corr(S_1, S_2)$. There after calculating hamming distance for each 3916 pairs. Hamming distance between two vectors of same length is number of positions at which corresponding values are different. In our case it gives number of times two stocks were in different clusters on a specific day. We then pick pairs which are together for p%(varying from 90% to 50%) of times and marked an edge between them. This way network within the stocks are built up. Figure 2 gives the prominent subgraphs.

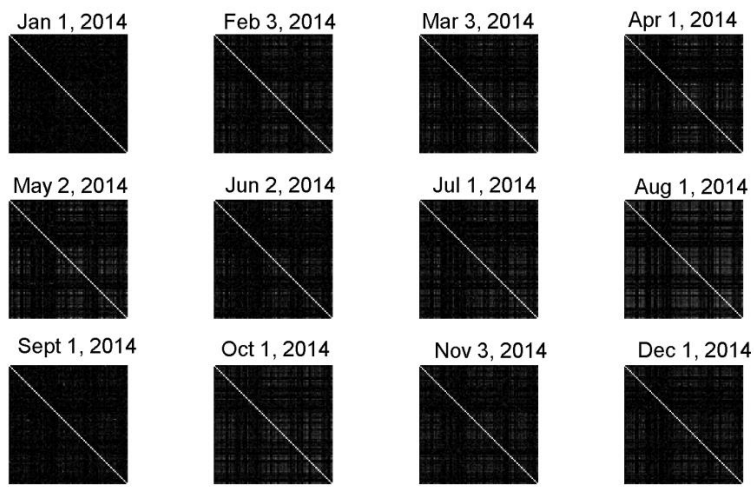

Figure 1: Correlation coefficient matrix for the first trading day of each month in form of an image. Correlation coefficient between each pair represented by grayscale image, ranging from black to white, minimum to maximum.

| Day | correlation coefficient | | no of pairs with correlation coefficient > 0.5 | no of pairs with correlation coefficient > 0.6 |
| --- | --- | --- | --- | --- |
| | max | min | | |
| Jan 1, 2014 | 0.2796 | -0.1351 | 0 | 0 |
| Feb 3, 2014 | 0.3913 | -0.1845 | 0 | 0 |
| Mar 3, 2014 | 0.4828 | -0.2005 | 0 | 0 |
| Apr 1, 2014 | 0.6639 | -0.153 | 10 | 1 |
| May 2, 2014 | 0.5162 | -0.1552 | 2 | 0 |
| Jun 2, 2014 | 0.4596 | -0.1532 | 0 | 0 |
| Jul 1, 2014 | 0.4029 | -0.1196 | 0 | 0 |
| Aug 1, 2014 | 0.4951 | -0.1489 | 0 | 0 |
| Sept 1, 2014 | 0.4045 | -0.1535 | 0 | 0 |
| Oct 1, 2014 | 0.3941 | -0.1441 | 0 | 0 |
| Nov 3, 2014 | 0.3662 | -0.1798 | 0 | 0 |
| Dec 1, 2014 | 0.381 | -0.1339 | 0 | 0 |

Table 3: Summary of the correlation coefficient obtained from each of 3916 pairs corresponding to first trading day of each month.

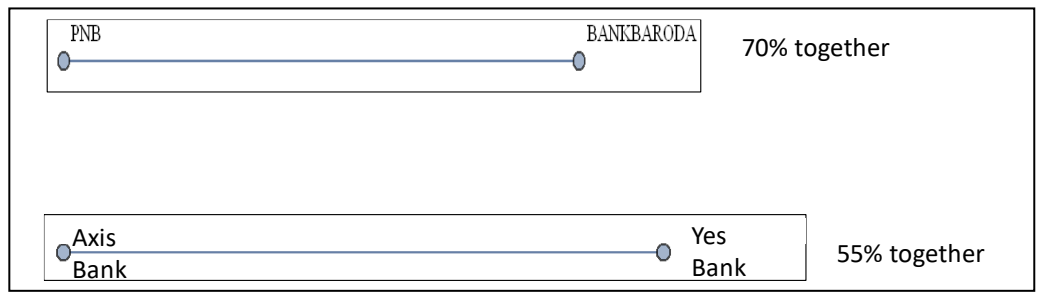

Figure 2:
Networks found in case of correlation coefficient method for hamming distance (a) <30% i.e. at least 70% of the times stocks are together, similarly (b) <45%

It is quite evident that PNB and Bank of Baroda, two nationalized banks, are seen to be moving hand in hand quite a number of days, 162 out of 229 days ~ 71% of times. Similar is the case with Axis Bank and Yes Bank, 131 out of 229 ~ 57%. Though the numbers are not so impressive but still one can relate the occurrence of these pairs in same clusters with the sector they come from.

We then further investigate this by running FPCA and KPCA with Gaussian Kernel with $\sigma = 1$, on the data for each 229 working days. Principal components that explained variability of at least 75%, varying till 92%, are picked and clustering procedure is applied to it. For each 229 days, we then use k-means clustering to put all the 89 stocks in various clusters. Again hamming distance is used to form the graphs in similar way. Figure 3 and figure 4 gives the prominent sub graphs for different p%.

All the programming is done in Matlab and R. Specifically for FPCA, we have used Matlab package fdaM "http://www.psych.mcgill.ca/misc/fda/downloads/FDAfuns/Matlab/ ".

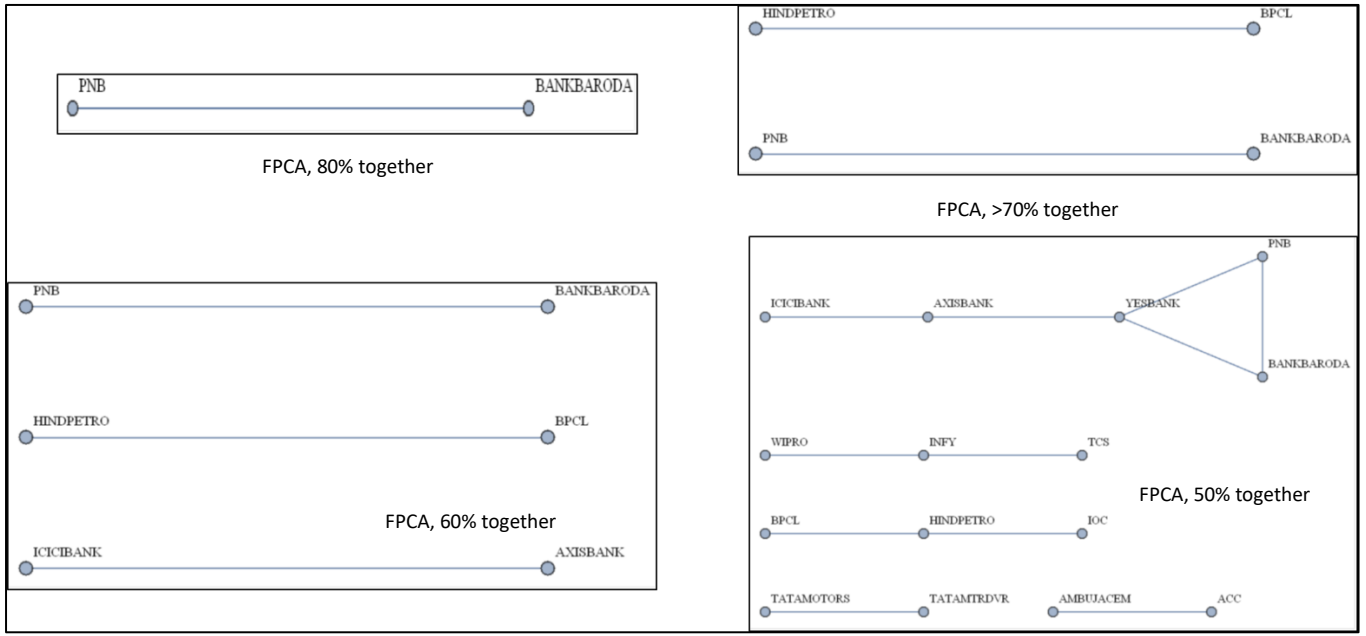

Figure 3: Networks found in case of FPCA for hamming distance (a) <20% i.e. more than 80% of the times stocks are together, similarly (b) <30% (c) <40% (d) <50%

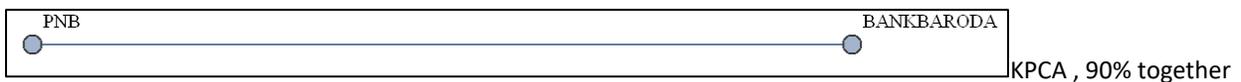

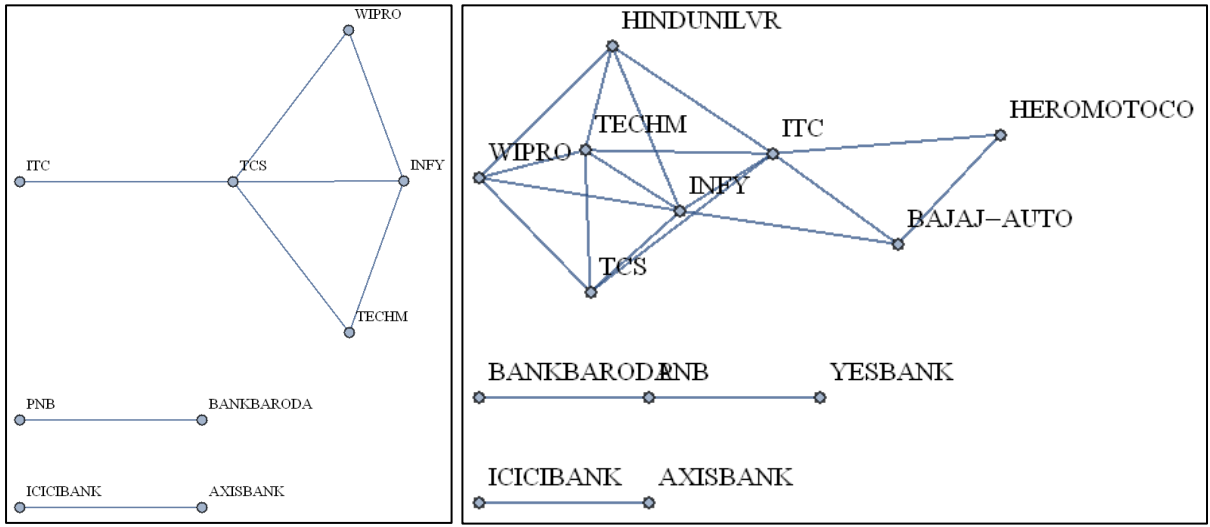

KPCA, 85% together　　　　　　　　　　　　KPCA, 80% together

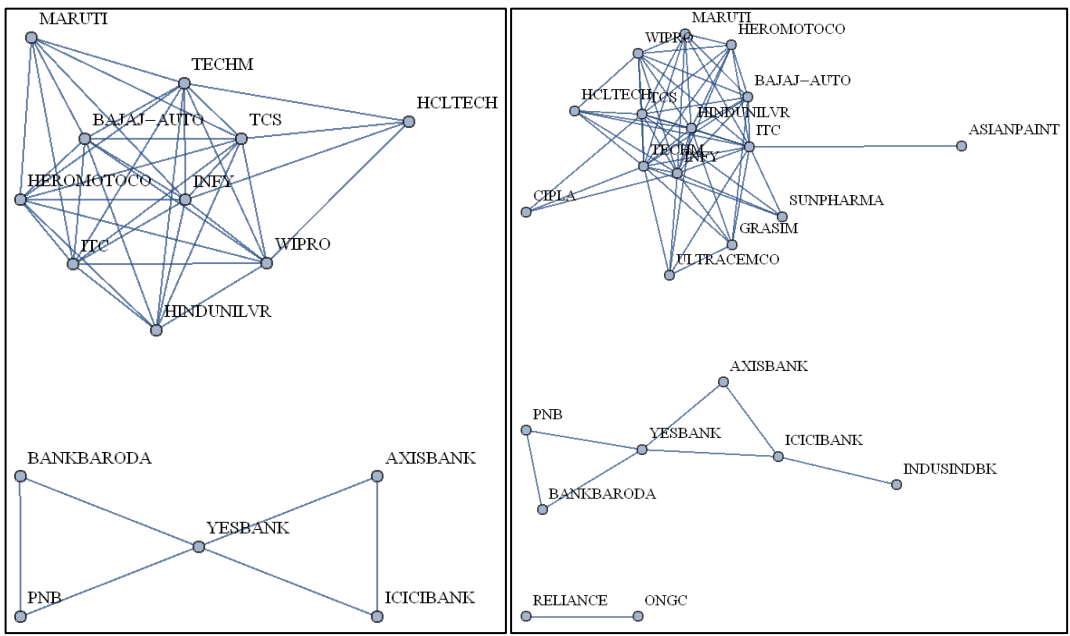

KPCA, 75% together　　　　　　　　　　　　KPCA, 70% together

Figure 4: Networks found in case of KPCA for hamming distance (a) <10% i.e. more than 90% of the times stocks are together, similarly (b) <15% (c) <20% (d) <25% (d) <30%

P values corresponding to the strength of the relationship is also calculated in case of all three methods. We performed hypothesis testing to test proportion of times the two stocks were together in the same group. Table 4 summarizes the p-values obtained following KPCA method corresponding to one tailed test, proportion of times two stocks are together 80% of times against that they are together more than 80% of times. Table 5 compares the p-values from different methods.

| S1 | S2 | No. of times together in a same cluster KPCA | z-statistic KPCA | p value KPCA |
|---|---|---|---|---|
| PNB | Bank of Baroda | 213 | 4.923098573 | 4.25923E-07 |
| ICICI | Axis Bank | 202 | 3.105847422 | 0.000948673 |
| TCS | Infy | 205 | 3.601461372 | 0.000158217 |
| TCS | Wipro | 198 | 2.445028822 | 0.007242028 |
| TCS | Techm | 198 | 2.445028822 | 0.007242028 |
| Infy | Wipro | 198 | 2.445028822 | 0.007242028 |
| Infy | Techm | 202 | 3.105847422 | 0.000948673 |

Table 4: Hypothesis testing for 7 strongest pairs considering KPCA. Test: one tailed test, proportion of times two stocks are together 80% of times against that they are together more than 80% of times.

| S1 | S2 | Rank correlation-coefficient | | FPCA | | KPCA | |
|---|---|---|---|---|---|---|---|
| | | no.of days pair was in same cluster | pvalue | no.of days pair was in same cluster | pvalue | no.of days pair was in same cluster | pvalue |
| PNB | Bank of Baroda | 162 | 4.03E-01 | 187 | 5.901E-05 | 213 | 1.49E-14 |
| ICICI | Axis Bank | 125 | 1 | 150 | 9.31E-01 | 202 | 9.09E-10 |
| TCS | Infy | 47 | 1 | 127 | 1 | 205 | 5.75E-11 |
| TCS | Wipro | 30 | 1 | 107 | 1 | 198 | 2.72E-08 |
| TCS | Techm | 32 | 1 | 101 | 1 | 198 | 2.72E-08 |
| Infy | Wipro | 28 | 1 | 117 | 1 | 198 | 2.72E-08 |
| Infy | Techm | 31 | 1 | 94 | 1 | 202 | 9.09E-10 |
| Yes Bank | Axis Bank | 131 | 1 | 119 | 1 | 177 | 8.02E-03 |
| HINDPETRO | BPCL | 123 | 1 | 175 | 1.70E-02 | 145 | 9.86E-01 |
| TATAMTRDVR | TATAMOTORS | 114 | 1 | 137 | 1 | 81 | 1 |

Table 5: Hypothesis testing for 10 prominent pairs, with sample size 229. Test: one tailed test, proportion of times two stocks are together 70% of times against that they are together more than 70% of times.

Keeping in mind that 2014 was a special year, General Elections were held in India during the month of April and May, it looked reasonable to look into this time period separately as well. We decided to pick data starting from March 2014, as promotional rallies were been held, extending till May 2014. With this set of data, same steps are repeated using all three different methods and the key features of this analysis are summarized as follows:

- The same networks showed up and this time they were seen to be more tightly bonded. Figure 5 gives the percentage a pair was found to be together in the same cluster. Also, p-values corresponding to top 10 pairs are shown in table 6.
- Network corresponding to the banking sector again emerged to be a standalone network as seen in previous analysis.

- IT network however was seen to be interacting with stocks from different sectors.

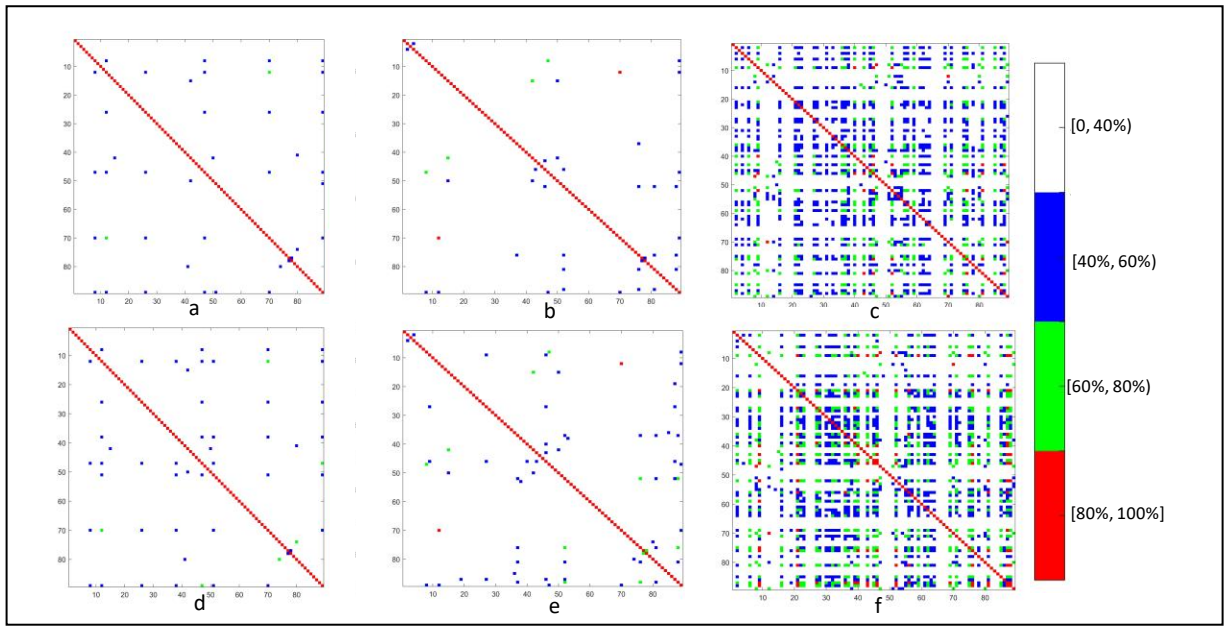

Figure 5: All 89 stocks are listed as rows and columns, and for each pair a colour represents the percentage($p$) the pair were together in the same cluster. Colour scheme: Red: $p \geq 80\%$, Green: $60\% \leq p < 80\%$, Blue: $40\% \leq p < 60\%$ and White: $p < 40\%$. (a) rank correlation, 2014 data (b) FPCA, 2014 data (c) KPCA, 2014 data (d) rank correlation, Mar 2014-May 2014 data (e) FPCA, Mar 2014-May 2014 data (f) KPCA, Mar 2014-May 2014 data

| 1 | S2 | Rank correlation-coefficient | | FPCA | | KPCA | |
|---|---|---|---|---|---|---|---|
| | | no.of days pair was in same cluster | pvalue | no.of days pair was in same cluster | pvalue | no.of days pair was in same cluster | pvalue |
| PNB | Bank of Baroda | 35 | 1.84E-01 | 38 | 3.10E-02 | 45 | 1.91E-05 |
| ICICI | Axis Bank | 22 | 1 | 35 | 1.84E-01 | 39 | 1.43E-02 |
| TCS | Infy | 11 | 1 | 33 | 3.98E-01 | 43 | 2.56E-04 |
| TCS | Wipro | 10 | 1 | 28 | 9.12E-01 | 43 | 2.56E-04 |
| TCS | Techm | 6 | 1 | 24 | 9.96E-01 | 40 | 6.04E-03 |
| Infy | Wipro | 11 | 1 | 29 | 8.48E-01 | 42 | 8.08E-04 |
| Infy | Techm | 7 | 1 | 22 | 9.99E-01 | 42 | 8.08E-04 |
| Yes Bank | Axis Bank | 23 | 1 | 20 | 1.00E+00 | 33 | 3.98E-01 |
| HINDPETRO | BPCL | 24 | 1 | 34 | 2.81E-01 | 32 | 5.26E-01 |
| TATAMTRDVR | TATAMOTORS | 25 | 1 | 30 | 7.60E-01 | 17 | 1.00E+00 |

Table 6: Hypothesis testing for 10 prominent pairs, with sample size 46, March 2014- May 2014. Test: one tailed test, proportion of times two stocks are together 70% of times against that they are together more than 70% of times

V. Conclusion

The aim of this paper is to study interactions between the stocks at tick by tick level for which we picked 30 second as our tick size and studied behavior of 89 stocks out of 100 stocks listed in CNX100 for the year 2014. Where the correlation coefficient actually fails to give us a detailed picture, KPCA with Gaussian kernel gives an insight to this level of analysis with higher power. Some of the sectors are seen to be in tight association and moving together. Stocks from the banking sector and the IT sector are seen to form a tight knitted networks respectively. Some of the stocks present under automobile industry are seen to be interacting with the IT hub while energy sector also emerged to be an independent network by itself. The knowledge of these interactions at this level will certainly be useful to our intraday traders while deciding their portfolios on a day to day basis.

Since IT and financial services sector emerged out to be tight knitted networks in our analysis, we plan to study these networks in depth in our future work. We aim to fit a multivariate stochastic model to each of these networks at the high frequency level. Fitting the right model will certainly help us to improve predictions of risk quantifiers like VaR(Value at risk) at the individual stock levels and thus also at portfolio level. The knowledge of estimated VaR can be then used by the brokering firms to set margin money for intraday stock traders.